\documentstyle[aps,prl,epsf,floats]{revtex}
\begin{document}
\twocolumn[\hsize\textwidth\columnwidth\hsize\csname@twocolumnfalse%
\endcsname

\title{{\it Ab initio} calculation of resonant X-ray scattering in Manganites }
\author { Patrizia Benedetti,$^1$ Jeroen van den Brink,$^{2,3}$  Eva Pavarini,$^{2,4}$ 
Assunta Vigliante,$^1$ and Peter Wochner$^1$}

\address{ $^1$Max-Planck-Institut f\"ur Metallforschung,
          Heisenbergstrasse 1, D-70569 Stuttgart, Germany }
\address{ $^2$Max-Planck-Institut f\"ur Festk\"orperforschung,
          Heisenbergstrasse 1, D-70569 Stuttgart, Germany }
\address{ $^3$Faculty of Applied Physics and MESA$^+$ Institute,
            University of Twente, Box 217, 7500 AE Enschede, The Netherlands}
\address{ $^4$INFM, Dipartimento di Fisca A.Volta, Via Bassi 6. I-27100 Pavia}

\date{\today}

\maketitle

\begin{abstract}
We study the origin of the resonant x-ray signal in manganites and
generalize the resonant cross-section to the band structure framework.  
With {\it ab initio} LSDA and LSDA+U calculations we determine the resonant x-ray 
spectrum of LaMnO$_3$.
The calculated spectrum and azimuthal angle dependence at the Mn $K$-edge 
reproduce the measured data without adjustable parameters.
The intensity of this signal is directly related to the orthorhombicity of the lattice. 
We also predict a resonant x-ray signal at the La $L$-edge, caused by the 
tilting of the MnO$_6$ octahedra. 
This shows that the resonant x-ray signal in the hard x-ray regime can be understood in 
terms of the band structure of a material and is sensitive to the fine details of 
crystal structure.
\end{abstract}

\pacs{PACS numbers}]

In strongly correlated $3d$ transition metal oxides (TMO's) orbital degeneracy and 
orbital order have a large effect on magnetic, structural and electronic properties.
The discovery of a Colossal Magneto-Resistance in perovskite manganites \cite{Jin94} 
has intensified experimental and theoretical efforts to understand and quantify the 
repercussions of orbital degeneracy on the physical properties of TMO's.
An orbitally
degenerate ground state is unstable and different mechanisms have been proposed 
for the lifting of the orbital degeneracy of the $3d$ electrons.
One possible mechanism was implicitly provided by Goodenough and Kanamori \cite{Gooden63},
who realized that superexchange interactions, which are due to the strong Coulomb
interactions between electrons, depend on the orbitals the spins 
actually occupy. This implies that magnetic and orbital order are directly related to each other,
and ordering of the orbitals, induced by the magnetic exchange, can lift the orbital 
degeneracy \cite{KK73}. 
On the other hand the lifting of the orbital degeneracy can also be induced
by lattice distortions: this is the well known Jahn-Teller
effect~\cite{Jahn37}.
In order to infer the actual orbital order in TMO's one can therefore make use of indirect
information coming from the magnetic and  (Jahn-Teller distorted) crystal structure. 
For instance in  vanadates (V$_2$O$_3$
\cite{Castellani78}, LiVO$_2$ \cite{Pen97}), cuprates 
(KCuF$_3$ \cite{KK73}) and manganites (LaMnO$_3$ \cite{Jirac85,Gooden63}) the orbital
order can be traced in this way.
The orbital structure is very important as it strongly
affects the low energy and low temperature behavior of these systems.
The description of the metal-insulator transition in V$_2$O$_3$ in terms of a spin 1/2 
Hubbard model \cite{McWhan69}, the magnon dispersion in (doped) manganites \cite{Moussa96} 
and charge-order in half-doped manganites \cite{Brink99} are all related to the 
underlying orbital structure of the material.

X-ray techniques, which could in principle serve as an independent direct probe of the
orbital order of the $d$ electrons, are under normal conditions not very sensitive to the 
distribution of valence electrons.
Nevertheless, it was recently demonstrated on perovskite manganites that the use of hard 
x-rays corresponding to the Mn $K$-edge, greatly enhances the sensitivity to the valence 
electron distribution and is related to the orbital 
order \cite{Murakami98}. This type of experiment is now widely used to probe
orbital and charge order in various TMO's~\cite{Zimmermann99}.
The question is, however, where the sensitivity to the valence electrons comes from.
Essentially three models have been proposed. Fabrizio {\it et al.}
\cite{Fabrizio98} suggest that for the manganites an $1s$ 
electron, excited at the $K$-edge resonance, makes an optically forbidden quadrupole
transition into the $3d$ valence band and directly probes the empty and occupied 
Mn $e_g$ orbitals. Ishihara {\it et al.} \cite{Ishihara98} argue that the 
transition is of dipolar character, into the Mn $4p$ states. 
The sensitivity to the  $3d$ states is then caused by the Coulomb repulsion between a $4p$ 
and $3d$ electron, which depends on the symmetry of the orbital that is occupied.
Finally, Elfimov {\it et al.} \cite{Elfimov99} suggest that the transition is dipolar, 
but that the signal is mainly due to the in-equivalence of the $4p$ bands due to the Jahn-Teller
distortion and therefore not directly sensitive to the Mn $3d$ states. 

The aim of this paper is to clarify which physical effects contribute to
the resonant x-ray scattering (RXS) signal in manganites.
In order to do this, it is necessary to make firm theoretical predictions for the 
RXS signal within a chosen model on a specific compound. 
We adopt the model of Elfimov {\it et al.} and generalize the resonant
cross-section to the band structure picture. By means of {\it ab initio} LSDA
and LSDA+U calculations for LaMnO$_3$ we determine the RXS spectrum and 
show that the calculated spectrum at the Mn-edge agrees in detail with 
experiment and reproduces the measured data without adjustable parameters.
This implies that the Jahn-Teller distortion is the main cause of the RXS signal. 
Via a tight-binding description of the bandstructure, we show that the intensity of the 
RXS signal is directly related to the orthorhombicity of the lattice. 
We also predict a resonant signal at the La $L$-edge, which was recently
observed in experiment~\cite{Gibbs}. The presence of a signal on the La-edge is a consequence
of the tilting of the MnO$_6$ octahedra and, to a lesser extent, of the Jahn-Teller distortion 
of the MnO$_6$ octahedra in the crystal and is independent of the electronic configuration
of the Mn $3d$ levels. 
This shows that the RXS signal is sensitive to the fine details of the crystal structure 
of a material.

\begin{figure}
\epsfxsize=85mm
\centerline{\epsffile{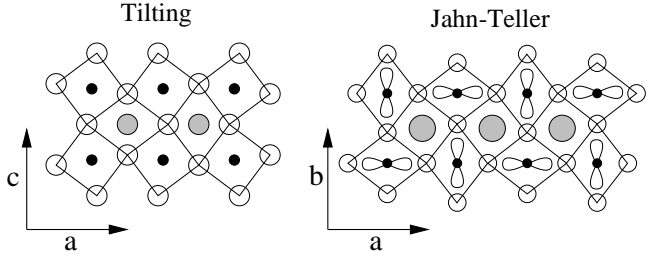}}
\vspace{1em}
\caption{Schematic representation of the crystal distortions of LaMnO$_3$.
Manganese atoms are denoted by the small filled circles, the oxygen by the 
large open circles and the lanthanum by the hatched circles. Left: projection 
on the $ac$ plane of the tilted octahedra. Right: projection on the $ab$ plane
of the Jahn-Teller distorted octahedra with the occupied $3x^2-r^2$ and $3y^2-r^2$
orbitals.}
\label{fig:JT}
\end{figure}

LaMnO$_3$ is a prototypical orbital ordered compound. 
The Mn$^{3+}$ ion is in a high spin $d^4$ configuration, where three electrons are in the $t_{2g}$
and one electron in the $e_g$ orbitals.
The orthorhombic $Pbnm$ crystal
structure can be considered as a cubic perovskite with two types of distortions:
first a tilting of the MnO$_6$ octahedra, so that the Mn-O-Mn angles become less
then $180^o$, and second a cooperative Jahn-Teller distortion (shortening of four Mn-O bonds
and elongation of the other two, see Fig.~\ref{fig:JT}). The twofold degeneracy of the
$e_g$ orbital, occupied by one electron, is lifted by the Jahn-Teller distortion.
Due to these distortions two neighboring Mn atoms are inequivalent: on one Mn-site mainly the
$3x^2-r^2$ orbital is occupied, while on the neighboring site the occupied orbital has mainly
$3y^2-r^2$ character. We show below that the in-equivalence of the local environment of 
neighboring Mn-atoms (and also of the La-atoms) gives rise to a large signal in a RXS experiment.

In order to calculate the RXS signal from first principles, we 
first present the formulae for the RXS cross section, that we use later on to obtain
the resonant spectra and azimuthal angle dependencies from the local density calculations.
We adopt the notation introduced by Ishihara and Maekawa~\cite{Ishihara98}
and define the structure factor
\begin{equation}
F_{i,\alpha \beta}^{{\bf k}-{\bf k}^{\prime}} = e^ { i ({\bf k}-{\bf k}^{\prime} ) {\bf R}_i } 
\sum_{b,{\bf q}}
\frac{ \langle 0 |P_{i,\alpha}|\psi{_{\bf q}}{^b} \rangle \langle \psi{_{\bf q}}{^b} |P_{i,\beta}| 0 \rangle }
{\omega-E_{\bf q}^b+E_0+ i \eta},
\end{equation}
where ${\bf k}$ (${\bf k}^{\prime}$) is the incident (outgoing) momentum of the light, $P_{i,\alpha}$
is the dipole operator with polarization $\alpha,\beta = x,y,z$ on site $i$ with lattice vector ${\bf R}_i$, 
$| 0 \rangle$ ($E_0$) is the groundstate wavefunction (energy), $|\psi{_{\bf q}}{^b} \rangle$ an 
excited state with momentum ${\bf q}$,
energy $E_{\bf q}^b$ and band-index $b$.
To be specific we consider the (3,0,0) reflection for the undoped manganite LaMnO$_3$ in
the experimental setup~\cite{Murakami98} at the Mn $K$-edge and assume that the incoming light is fully
$\sigma$ polarized. At the $K$-edge the relevant dipole operator is
\begin{equation}
P_{i,\alpha} = p^{\dagger}_{i,\alpha} s_i + s^{\dagger}_i p_{i,\alpha}, 
\end{equation}
where $s_i$ creates a $1s$ core hole on the Mn and $p^{\dagger}_{i,\alpha}$ creates an electron in the
Mn 4p shell with $\alpha = x,y,z$ symmetry.
A detailed derivation of the cross-section in the $\sigma \rightarrow \pi$ channel, where the 
polarization state of the photon is rotated by $\pi /2$ on scattering, shows that the
intensity is given by
\begin{equation}
I_{\sigma \rightarrow \pi} = const \ | \sin \theta A_{21} (\phi) - \cos \theta A_{31} (\phi)|^2,
\end{equation}
where $\theta$ is the scattering angle, which is fixed in experiment at approximately $\pi/6$, and $\phi$ the azimuthal angle which describes the rotation around the scattering vector,
\begin{eqnarray}
&A&_{21} (\phi) = \cos^2 \phi (F_{xz}+F_{yz}) - \sin^2 \phi (F_{zx}+F_{zy})  \nonumber \\
&-& \sin \phi \cos \phi (F_{xx}+F_{yy}+F_{xy}+F_{yx}-2F_{zz})/\sqrt{2},
\end{eqnarray}
and
\begin{eqnarray}
A_{31} (\phi) &=& \cos \phi (F_{xz}-F_{yz}) \nonumber \\
&-& \sin \phi (F_{xx}-F_{yy}+F_{xy}-F_{yx})/\sqrt{2}. 
\label{eq:A31}
\end{eqnarray}
In LaMnO$_3$ the orbital ordered state consists of two Mn sublattices, where on sublattice A (B) presumably
mostly the $3x^2-r^2$ ($3y^2-r^2$) orbital is occupied, see Fig~\ref{fig:JT}. 
The RXS cross section is exactly
sensitive to the difference between the two sublattices, as
\begin{eqnarray}
F_{\alpha \beta} = \sum_{ i \in A} F_{i,\alpha \beta}^{\bf 0}-\sum_{ i \in B} F_{i,\alpha \beta}^{\bf 0}.
\end{eqnarray}
From this expression it is easy to deduce the physical interpretation of RXS cross-section in 
the band-picture. $F_{xx}$, for instance, is simply proportional to the difference of the Mn $4p_x$ projected 
density of states (DOS) on sublattice A and the Mn $4p_x$ projected DOS on sublattice B. The RXS
intensity is proportional to the square of this difference. 
Note that off-diagonal elements of the structure factor tensor are non-zero due
to the tilting of the octahedra.

The electronic structure is calculated with the LMTO ASA LSDA method~\cite{Andersen84} and with 
the LSDA+U method~\cite{Anisimov91}, that incorporates a mean-field treatment 
of the strong correlations of the $e_g$ electrons. All the {\it ab-initio} calculations are based on 
the Stuttgart TBLMTO47 code.
%
%
We perform the calculations on a set of different (virtual) crystal structures
for LaMnO$_3$. First of all the RXS signal is calculated for the real lattice structure,
including the tilting (T) and Jahn-Teller (JT) distortion, both with and without the on-site Coulomb 
interaction $U$.
In order to study the effect of tilting and Jahn-Teller distortion separately, we do the same calculations
for LaMnO$_3$ in the not Jahn-Teller distorted (but tilted) Pr$_{1/2}$Sr$_{1/2}$MnO$_3$ structure and
for a virtual crystal structure that is Jahn-Teller distorted, but not tilted.

\begin{figure}
\epsfxsize=80mm
\centerline{\epsffile{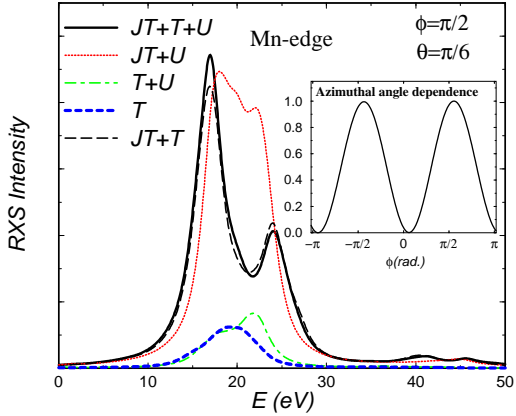}}
\caption{Resonant x-ray scattering intensity at the Mn-edge from LSDA(+U) calculations for different 
crystal structures. JT+T: experimental Jahn-Teller distorted and tilted $Pbnm$ crystal structure of
LaMnO$_3$, JT (T): virtual crystal structure with only the Jahn-Teller distortion (tilting of the
MnO$_6$ octahedra). The Fermi-level is chosen as the zero of energy. 
Inset: azimuthal angle dependence of the main edge from the JT+T+U calculation
normalized to the maximum intensity.}
\label{fig:I_Mn}
\end{figure}

We first present the results for the Mn $K$-edge in Fig.~\ref{fig:I_Mn}, calculated with
Eq.(1-6), where the core hole potential is neglected, and the core-hole life-time leads to a 
broadening of the signal with a  half width at half maximum of 1.2 eV~\cite{Elfimov99}. 
In LSDA+U calculations we use $U$=8 eV~\cite{Satpathy96}.
The spectrum for the full
calculation (JT+T+U) shows that at ~10 eV above the absorption edge
there is a strong satellite (experimentally ~12 eV~\cite{Murakami98,Wochner_P}) 
and weak satellites 23 and 27 eV above the edge, that are also found experimentally at the same 
energies\cite{Wochner_P}. The calculation for the JT+T system (U=0) 
%
%
agrees with the result of Takahashi {\it et al.}~\cite{Taka99} and
is very
similar to the JT+T+U result, with a slight shift of spectral weight from the main peak to the
satellite. Electron correlations, that are responsible for the orbital dependent
superexchange and a possible driving force for orbital order~\cite{KK73}, therefore make a negligible
contribution to the RXS intensity at the Mn $K$-edge, as is suggested by Elfimov 
{\it et al.}~\cite{Elfimov99} and in agreement with multiple scattering cluster
calculations~\cite{Benfatto99}.
 The azimuthal angle dependence at the main peak shows that the signal 
follows closely a $\sin^2 \phi$ behavior, in accordance with experiment~\cite{Murakami98}, 
with a phase factor that is slightly shifted by $0.059 \pi$. 
This indicates that the main contribution is due
to the term $F_{xx}-F_{yy}$ in equation (\ref{eq:A31}), which comes from the difference 
in $p_x$ ($p_y$) projected density of states on the two sublattices and is obviously connected
to the JT distortion. This assignment is supported by the calculation for the system without
Jahn-Teller distortion shown in Fig.\ref{fig:I_Mn}. In the T+U system the signal is broad,
featureless, an order of magnitude weaker and almost entirely due to the tilting.
From the calculation for the JT+U system we conclude that the tilting of the octahedra is
responsible for the observed satellites, but has only a moderate effect on the total intensity
of the RXS signal at the Mn $K$-edge. 

\begin{figure}
\epsfxsize=70mm
\centerline{\epsffile{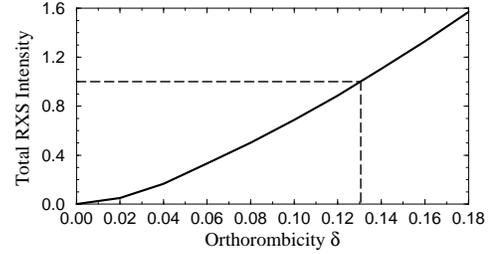}}
\vspace{1em}
\caption{Total resonant x-ray intensity as a function of the Jahn-Teller distortion strength
calculated within a tight-binding framework, normalized to the total intensity at the experimental
value for the orthorhombicity (defined in the text) $\delta= 0.130$ (dashed line).}
\label{fig:I_JT}
\end{figure}

An interesting question is how the RXS signal actually depends on the strength of the Jahn-Teller 
distortion. From the LSDA calculation we obtain the 
parameters for a tight binding (TB) description of the electronic structure for a lattice without tilting 
and JT distortion, in accordance with previously published results~\cite{Picket98}.
The RXS intensity for the undistorted lattice and $U=0$ is of course vanishing because all 
neighboring Mn atoms are equivalent. A JT distortion modifies the TB parameters, whose distance
dependence is given by Harrison~\cite{Harrison80} and allows the resonant scattering at (3,0,0). 
In Fig.\ref{fig:I_JT} the total RXS intensity is shown as a function of the orthorhombicity.
The JT distortion gives rise to a long ($l$) and short ($s$) manganese-oxygen bondlength in 
the $ab$ plane
and we define the orthorhombicity as $\delta = (l-s)/(l+s)$ so that for the cubic perovskite
$\delta=0$. The experimental
value for the orthorhombicity at low temperature is $\delta= 0.130$~\cite{Moussa96}.
The resonant intensity smoothly 
depends on $\delta$ and is approximately proportional to $\delta^{3/2}$. 
This supports the suggestion that the temperature dependence of the RXS is closely related to 
the temperature dependence of the lattice parameters~\cite{Wochner_P}.

\begin{figure}
\epsfxsize=60mm
\centerline{\epsffile{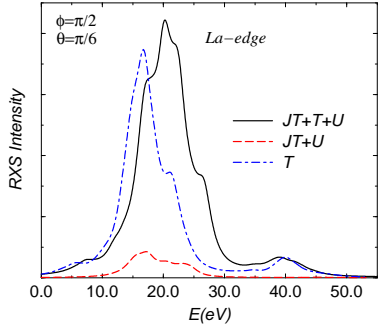}}
\caption{Resonant x-ray scattering intensity at the La-edge from LSDA(+U) calculations for different 
crystal structures. JT+T: experimental Jahn-Teller distorted and tilted crystal structure, 
JT (T): virtual crystal structure with only the Jahn-Teller distortion (tilting of the MnO$_6$ octahedra).}
\label{fig:I_La}
\end{figure}

Finally we present the results for the RXS signal at the La $L$-edge. Fig.~{\ref{fig:JT}} shows that
due to the lattice distortions also neighboring La-atoms have locally different surroundings.
We calculate the RXS intensity for the (3,0,0) reflection in the $\sigma \rightarrow \pi$ channel,
for different (virtual) lattice structures, as we did for the Mn-edge. We find a signal at the
La-edge, shown in Fig.~\ref{fig:I_La}, that is mainly due to the tilting. From the JT+U calculation
we see that the combined contribution of the other two symmetry breaking effects is small.
For the crystal structure that only has the tilting distortion we find a signal that is comparable
to the JT+T+U calculation both in intensity and shape.
This result is not unexpected as the tilting of the MnO$_6$ octahedra modifies the local environment
of La atoms: the apical oxygen atoms are displaced towards the La atoms on one sublattice and away from
them on the other sublattice so that neighboring lanthanum atoms have inequivalent oxygen surroundings. 

We determined, in conclusion,
with {\it ab initio} LSDA and LSDA+U calculations the resonant x-ray spectrum (RXS) of LaMnO$_3$.
The calculated spectrum and azimuthal angle dependence at the Mn-edge reproduce the measured 
data without adjustable parameters. Spectra for different (virtual) crystal structures 
show that the RXS intensity at the Mn $K$-edge is mainly due to the Jahn-Teller distortion 
%
%
\cite{Elfimov99}
and 
that the 
tilting of the MnO$_6$ octahedra causes the strong satellite structure in the spectrum.
We find that the intensity of the resonant x-ray signal at the Mn-edge is directly related 
to the orthorhombicity of the lattice. 
We also predict a RXS signal at the La $L$-edge, which is a 
consequence of the tilting of the MnO$_6$ octahedra in the crystal. 
This shows that the resonant signal at different edges in the hard x-ray regime can be understood
and predicted by calculations in the band-structure framework and 
%
%
%
%
is sensitive to
the fine details of lattice distortions, but does not 
%
%
{\it directly} 
probe 
the orbital order of the $3d$ states. 
Orbital order, of large enough periodicity to satisfy Bragg conditions, 
might however be visible in resonant 
experiments with soft x-rays, where direct transitions into the $3d$ states are allowed.

We thank I. Elfimov, G. Sawatzky and P.J. Kelly for useful discussions.


\end{document}